\documentclass[11pt]{article}

\usepackage{a4}
\usepackage{latexsym,amssymb}
\usepackage{citesort}

\usepackage{url}
\usepackage{ntheorem}

\newcommand{\newF}{\lambda}

\newcommand{\Hess}{\mbox{\rm Hess}}
\newcommand{\divv}{\mbox{\rm div}}

\newcommand{\mcO}{{\mycal O}}
\newcommand{\mcF}{{\mycal F}}

\newcommand{\myuu}{u}

\newcommand{\Ric}{{\mathrm{Ric}}}

{\catcode `\@=11 \global\let\AddToReset=\@addtoreset}
\AddToReset{equation}{section}

\newcommand{\tr}{\mbox{tr}}

\DeclareFontFamily{OT1}{rsfs}{}
\DeclareFontShape{OT1}{rsfs}{m}{n}{ <-7> rsfs5 <7-10> rsfs7 <10-> rsfs10}{}
\DeclareMathAlphabet{\mycal}{OT1}{rsfs}{m}{n}
\def\scri{{\mycal I}}%
\newcommand{\bel}[1]{\begin{equation}\label{#1}}
\newcommand{\bea}{\begin{eqnarray}}
\newcommand{\beaa}{\begin{eqnarray*}}
\newcommand{\bean}{\begin{eqnarray}\nonumber}
\newcommand{\beal}[1]{\begin{eqnarray}\label{#1}}
\newcommand{\eea}{\end{eqnarray}}
\newcommand{\eeaa}{\end{eqnarray*}}

\newcommand{\eeal}[1]{\label{#1}\end{eqnarray}}
\newcommand{\bed}{\begin{deqarr}}
\newcommand{\eed}{\end{deqarr}}
\newcommand{\bedl}[1]{\begin{deqarr}\label{#1}}
\newcommand{\eedl}[2]{\arrlabel{#1}\label{#2}\end{deqarr}}

\newcommand{\beq}{\begin{equation}}
\newcommand{\eeq}{\end{equation}}
\newcommand{\beqa}{\begin{eqnarray}}
\newcommand{\eeqa}{\end{eqnarray}}
\newcommand{\cref}[1]{4\emph{\ref{#1})}}

\newcommand{\eq}[1]{(\ref{#1})}

 \def\scri{\hbox{${\cal J}$\kern -.645em {\raise
      .57ex\hbox{$\scriptscriptstyle (\ $}}}}

\newcommand{\be}{\begin{equation}}
\newcommand{\ee}{\end{equation}}

{\catcode `\@=11 \global\let\AddToReset=\@addtoreset}
\AddToReset{equation}{section}

\newcounter{mnotecount}[section]

\newcommand{\oldmnote}[1]{ \marginpar{\raggedright\tiny\em old mnote
  in the file here  (to be discarded as far as PC is   concerned)}}
\newcommand{\oldnote}[1]{}  
\newcounter{pcheckcount}[section]

\newcommand{\pcheck}[1]{}

\begin{document}

\title{The asymptotics of stationary electro-vacuum metrics in odd space-time dimensions}

\author{
Robert Beig\thanks{E--mail \protect\url{robert.beig@univie.ac.at}}
\\
Faculty of Physics\\
 Universit\"at Wien \\ \\
Piotr T.\ Chru\'sciel\thanks{ E-mail
    \protect\url{Piotr.Chrusciel@lmpt.univ-tours.fr}, URL
    \protect\url{ www.phys.univ-tours.fr/}$\sim$\protect\url{piotr}}
  \\ LMPT,
F\'ed\'eration Denis Poisson\\
Tours
\\}

\maketitle

\begin{abstract}
We show that, for stationary, asymptotically flat solutions of the
electro-vacuum Einstein equations, the Riemannian metric on the
quotient space of the timelike Killing vector is analytic at $i^0$,
for a large family of gauges, in odd space-time dimensions larger
than seven. The same is true in space-time dimension five for static
vacuum solutions with non-vanishing mass.
\end{abstract}

\section{Introduction}
\label{SI}

There is currently interest in asymptotically flat solutions of the
vacuum Einstein equations in higher
dimensions~\cite{EmparanReallReview,BCS}. It is thus natural to
enquire which part of our body of knowledge of $(3+1)$--dimensional
solutions carries over to $n+1$ dimensions.  In this note we study
that question for asymptotic expansions at spatial infinity of
stationary or static electro-vacuum metrics. In \emph{even space
dimensions} $n\ge 6$, we prove analyticity  at $i^0$ of the
metric%
\footnote{We emphasise that, in the \emph{non}--static case, this is
\emph{never} the metric induced on some spacelike slice of
space-time. On the other hand, it follows from our results below
that electro-vacuum initial data  on well chosen (e.g.\ maximal)
hypersurfaces in even space dimensions $n\ge 6$ will have an
asymptotic expansion in terms of powers of $1/r$ (thus, without log
terms).}
 on the distribution orthogonal to the timelike Killing vector, up to a conformal factor, as
well as of the remaining (rescaled) components of the metric in a family of natural geometric
gauges. The same result is established in space dimension $n=4$ for static vacuum metrics with
non-vanishing ADM mass.

It would be of interest to extend the \emph{static} $n=4$ result to
the \emph{stationary} case, left open by our work. Similarly it
would be of interest to clarify the situation for odd space
dimensions $n$.

\section{Static vacuum metrics}

We write the space-time metric in the form
$$
ds^2 = -e^{2\myuu }dt^2 + e^{-\frac{2\myuu}{n-2} } \tilde
g_{ij}dx^idx^j
 \;,
$$
where $\tilde g$ is an asymptotically flat Riemannian metric, with
$\partial_t \myuu  = \partial_t \tilde g_{ij}=0$. The vacuum
Einstein equations show that $\myuu $ is $\tilde g$--harmonic, with
$\tilde g$ satisfying the equation
\bel{Eins1} \tilde R_{ij} = \frac {n-1}{n-2} \partial_i \myuu
\partial _j \myuu
 \;,
 \ee
where $n$ is the space dimension, and $\tilde R_{ij}$ is the Ricci
tensor of $\tilde g$. We assume $n\ge 3$ throughout.

Methods known in principle%
 \footnote{ In harmonic coordinates the equations form a quasi-linear elliptic system with diagonal
  principal part, with principal symbol identical to that of
   the scalar Laplace operator. One can then use  the results in, e.g., \cite{ChAFT} together with a simple
 iterative argument to obtain the expansion. We show below that the stationary vacuum, or electro-vacuum, equations
 imply a system with identical properties in several gauges, which again leads to a similar expansion. This analysis holds in any dimension;
 the restrictions on dimensions imposed in our main results here arise from the requirement  of simple
 conformal properties of
 the equations.}
 show that, in harmonic space-time coordinates in the asymptotically flat region, and whatever
$n\ge 3$, both $\myuu $ and $\tilde g_{ij}$ have a full asymptotic expansion in terms of powers of
$\ln r$ and inverse powers of $r$. Solutions without $\ln r$ terms are of special interest, because
the associated space-times contain smoothly compactifiable hyperboloidal hypersurfaces.
In even space-time dimension initial data sets containing such
asymptotic regions, when close enough to Minkowskian data, lead to
asymptotically simple
space-times~\cite{friedrich,AndersonChruscielConformal}. It has been
shown by Beig and Simon that logarithmic terms can always be gotten
rid of by a change of coordinates when space dimension equals
three~\cite{BeigSimon,SimonBeig}.

{}From what has been said one can infer that the leading order
corrections in the metric can be written in a Schwarzschild form,
which in ``isotropic" coordinates reads
\bean
 g_m&=&  - \left(\frac{1-m/2|x|^{n-2}}{1+m/2|x|^{n-2}}\right)^2
dt^2+\left(1 + \frac{m}{2|x|^{n-2}}\right)^{\frac4{n-2}}
\left(\sum_{1=1}^n dx_i^2\right)
 \\
 & \approx &
- \Big(1-\frac {m}{\tilde r^{n-2}}\Big)^2 dt^2+\Big(1+\frac
{m}{\tilde
 r^{n-2}}\Big)^{\frac 2 {n-2}}\left(\sum_{1=1}^n dx_i^2\right)
 \;,
 \eeal{stschw}
 where $m$ is of course a constant, and $\tilde r=|x|$ is a radial coordinate
in the asymptotically flat region. This gives the asymptotic
expansion
\bel{Uexp} \myuu  = -\frac m {\tilde r^{n-2}} + O(\tilde r^{-n+1})
 \;,
 \ee
Further we have
\bel{gtildeasym} \tilde g_{ij}=\delta_{ij}+O(\tilde r^{1-n})
 \;.
\ee
 Equation~\eq{Uexp} shows that for $m\ne 0$ the function
\bel{omdef} \omega:= (\myuu^2) ^{\frac 1 {n-2}}
 \ee
behaves asymptotically as $\tilde r^{-2}$, and can therefore be used
as a conformal factor in the usual one-point compactification of the
asymptotic region. Indeed, assuming that $m\ne 0$ and setting
\bel{confmetdef} g_{ij}:= \omega^2 \tilde g_{ij}
 \;.
\ee
one obtains a $C^{n-2,1}$ metric\footnote{The differentiability
class near $i^0$ can be established by examining Taylor expansions
there.}
on the manifold obtained by adding a point (which we denote by
$i^0$) to the asymptotically Euclidean region.

{}From the fact that $\myuu$ is $\tilde g$--harmonic one finds
\bel{omegaeq} \Delta \omega = \mu\;,
 \ee
where the auxiliary function $\mu$ is defined as
\bel{mudef} \mu:= \frac n 2 \omega^{-1} g^{ij}\partial_i \omega
\partial_j \omega
 \;.
 \ee
Note that, in spite of the negative power of $\omega$, this function
can be extended by continuity to $i^0$, the extended function, still
denoted by $\mu$, being of $C^{n-2,1}$ differentiability class.

Let $L_{ij}$ be the Schouten tensor of $g_{ij}$,
\bel{Schoutendef}
 L_{ij} := \frac 1 {n-2} \Big( R_{ij} - \frac R
{2(n-1)} g_{ij}\Big)
 \;.
 \ee
Using tildes to denote the corresponding objects for the metric
$\tilde g$, {}from \eq{Eins1} one obtains
\bel{Sch1} \tilde L_{ij} = \frac 14 \omega^{n-4} \Big(
(n-1)\partial_i \omega \partial_j \omega - \frac 12
g_{ij}g^{k\ell}\partial_k \omega \partial_\ell \omega\Big)
 \;.
\ee
We see that for $n\ge 3$ the tensor $\tilde L_{ij}$ is bounded on
the one-point compactification at infinity, and for $n\geq4$ it is
as differentiable as $d\omega$ and the metric allow. This last
property is not true anymore for $n=3$, however the following
objects are well behaved:
\bel{Lcont} \tilde L_{ij}D^j\omega=\frac {2n-3} {4n} \omega^{n-3}
\mu D_i \omega
 \;,\qquad \tilde L_{i[j}D_{k]}\omega=-\frac {1} {4n} \omega^{n-3}
\mu g_{i[j}D_{k]} \omega
 \;.
 \ee

\section{Conformal rescalings}

We recall the well-known transformation law of the Schouten tensor
under the conformal rescaling \eq{confmetdef}
\bel{Schouttranslaw} L_{ij} = \tilde L_{ij} + \omega^{-1}
D_iD_j\omega - \frac 12 \omega^{-2} g_{ij}g^{k\ell}\partial_k \omega
\partial_\ell \omega
 \;;
 \ee
we emphasize that this holds whether or not $\omega$ is related to
$\tilde L$ as in \eq{Sch1}. Taking a trace of \eq{Schouttranslaw}
and using \eq{omegaeq} one finds
\bel{Req} R=
\frac {(n-1)(n-2)}{2n}
\omega^{n-3} \mu
 \;.
\ee

In our subsequent manipulations it is convenient to rewrite
\eq{Schouttranslaw} as an equation for $D_iD_j\omega$,
\bel{Schouttranslaw2} D_jD_i\omega = \omega(L_{ij} - \tilde L_{ij})
+ \frac 1n \mu g_{ij}
 \;.
 \ee
 Differentiating \eq{mudef} and using \eq{Schouttranslaw2} one
obtains
\bean D_i\mu &=&  -n (L_{ij}-\tilde L_{ij})D^j\omega
  \\
  & = & -n L_{ij}D^j\omega +\frac {2n-3} {4} \omega^{n-3}
\mu D_i \omega\;,
 \eeal{divlapmu}
which allows us to eliminate each derivative of $\mu$ in terms of
$\mu$,  $\omega$ and $d\omega$.

 Let $C_{ijk}$ denote the Cotton
tensor,
\bel{Cottondef} C_{ijk}:= D_kL_{ij}- D_j L_{ik}
 \;,
\ee
and let $C_{ijkl}$ be the Weyl tensor of $g$. Note the identity
 \bel{DC}
 D^i C_{ijkl} = (3 - n) C_{jkl} \;.
 \ee

Inserting $\tilde L_{ij}$ from \eq{Sch1} into the right-hand-side of
\eq{Schouttranslaw2}, one finds that this right-hand-side is
well-behaved at $\omega=0$, in the sense of being analytic in
$\omega$ and its first derivatives, for all $n\ge 3$ when a smooth
metric $g$ is given.  Applying $D_k$ to \eq{Schouttranslaw2} and
anti-symmetrising over $j$ and $k$ one obtains
\bel{Ceq}
 \omega C_{ijk} + C_{kji\ell}D^\ell\omega = \tilde L_{ijk}
 \;,
\ee
where
\bel{tLdef}
 \tilde L_{ijk} := 2D_{[k}(\omega\tilde L_{j]i}) + 2 g_{i[j}\tilde
 L_{k]\ell}D^\ell\omega\;.
\ee
Writing down the second term in \eq{tLdef} and using \eq{Lcont} and
again \eq{Schouttranslaw2}, we find that the terms with
$\omega^{n-4}$ drop out and there results
\bel{tLdef1}
 \tilde L_{ijk} =  \frac 12 \omega^{n-2}\Big(\underbrace{ -(n-1)D_{[j}\omega(L_{k]i}- \tilde
 L_{k]i}) + g_{i[j}(L_{k]\ell}- \tilde
 L_{k]\ell}) D^\ell\omega}\Big)
 \;. \phantom{xxxx}
\ee
Here $\tilde L_{ij}$ should be expressed in terms of $\omega$,
$d\omega$ and $\mu$ using \eq{Lcont}. It should be emphasized that
the underbraced expression is again well-behaved at $\omega=0$.

 Let $B_{ij}$ denote the Bach
tensor,
\bel{Bachdef} B_{ij}:= D^kC_{ijk}- L^{k\ell} C_{kji\ell}
 \;.
\ee
Applying $D^k$ to \eq{Ceq} and using \eq{DC} and
\eq{Schouttranslaw2} one obtains
\bea
 B_{ij} -(n-4)\omega^{-1}C_{ijk} D^k\omega
 & = &  \omega^{-1}
 D^k 
 \tilde L_{ijk}
 - C_{kji\ell} \tilde L^{k\ell}
 \;.
\eeal{mainid}
Note that the factor $\omega^{-1}$ in front of  the divergence
 $D^k 
 \tilde L_{ijk}
 $
 is
compensated  by $\omega^{n-2}$ in \eq{tLdef}, so that (in view of
\eq{Sch1}) for $n\ge 4$
the right-hand-side is a well-behaved function of the metric, of
$\omega$, and of their derivatives at zeros of $\omega$.
Alternatively we can, using \eq{Ceq}, rewrite \eq{mainid} as
 \begin{equation}\label{mainid1}
B_{ij} + (n - 4)\omega^{-2}C_{kijl} D^k\omega  D^l\omega =
\omega^{3-n} D^k(\omega^{n-4} \tilde L_{ijk}) - C_{kijl} \tilde
L^{kl}
 \;.
\end{equation}
Note that the right-hand-side of (\ref{mainid1}) is regular also for $n=3$.
 Recall, now, the identity
\bean B_{ij} & =&
 \Delta L_{ij} - D_i D_j(\tr L) + \mcF_{ij}
 \\
 &=& \frac 1 {n-2}\Delta R_{ij} - \frac 1 {2(n-1)} \Big(\frac 1 {n-2}
\Delta R \,g_{ij}  + D_i D_j R\Big)+ \mcF_{ij}
 \;,
\eeal{Bacheq}
where $\mcF_{ij}$ depends upon the metric and its derivatives up to
order two. We  eliminate the Ricci scalar terms using \eq{Req}. The
terms involving derivatives of $R$
%
%
 will introduce derivatives of $\mu$, which are handled by \eq{divlapmu}.

\subsection{Space-dimensions three and four}

 In dimension three the  term involving
$\omega^{-2}C_{kji\ell} L^{k\ell}$ on the left-hand-side of
\eq{mainid1} goes away because the Weyl tensor vanishes. In
dimension four its coefficient vanishes. In those dimensions one
therefore ends up with an equation of the form
\bel{Riccieq} \Delta R_{ij} = F_{ij}(n,\omega,d\omega,\partial^2
\omega,g,\partial g, \partial^2 g)
 \;.
\ee
with a tensor field $F_{ij}$ which is an analytic function in its
arguments on a suitable open set. Here we have used the expression
of $\mu$ as a function of the metric, $\partial g$, $\partial
\omega$ and $\partial^2 \omega$ which follows {}from \eq{omegaeq}.

We can calculate the laplacian of $\mu$ by taking a divergence of
\eq{divlapmu} and eliminating again the second derivatives of
$\omega$ in terms of $\mu$, and the first derivatives of $\mu$, as
before. This leads to a fourth-order equation for $\omega$ of the
form
\bel{mueq}
 \Delta^2 \omega = F(n,\omega,d\omega,\partial^2 \omega,g,\partial g, \Ric)
 \;,
\ee
with $F$ analytic on a suitable set, where $\Ric$ stands for the
Ricci tensor. Note that one should use the Bianchi identities to
eliminate the term involving the divergence of $L_{ij}$ which arises
in the process:
$$
D^jL_{ij} = \frac 1 {2(n-1)} D_i R
 \;.
$$

In harmonic coordinates, Equations~\eq{Riccieq}-\eq{mueq} can be
viewed as a system of equations of fourth order for the metric $g$
and the function $\omega$, with diagonal principal part $\Delta^2$.
The system is elliptic  so that usual bootstrap arguments show
smoothness of all fields. In fact the solutions are real-analytic
by~\cite{Morrey}, as we wished to show.

\subsection{Higher even dimensions}
 \label{Sstatn}

A natural generalisation of the Bach tensor in even dimensions $n\ge
6$ is the obstruction tensor $\mcO_{ij}$ of Fefferman and
Graham~\cite{FG,GrahamHirachi}. It is of the form
\bel{ObsTens} \mcO_{ij}  = \Delta^{\frac{n-4}{2}}[\Delta L_{ij} -
D_i D_j(\tr L)] + \mcF^n_{ij}\;,
\end{equation}
where $\mcF^n_{ij}$ is a tensor constructed out of the metric and
its derivatives up to order $n-2$. This leads us to expect that
further differentiations of the equations above  leads to a regular
expression for $\Delta^{\frac{n-2}{2}}B_{ij}$ in terms of $\omega$
and its derivatives up to order $n-3$. However, we have not been
able to conclude using this approach. Instead, we proceed as
in~\cite{CCL}:

In coordinates $x^i$ which are harmonic with respect to the metric
$\tilde g$, \eq{Eins1} and the harmonicity condition for $u$ lead to
a set of equations for $u$ and
$$
 f:=(\tilde g_{ij}-\delta_{ij})
$$
of the form
$$
\tilde g^{ij}\partial_i\partial_j f = F(f) (\partial f)^2 +
(\partial u)^2 \;, \quad
 \tilde g^{ij}\partial_i\partial_j u = 0
$$
(an explicit form of those equations with $u=1$ can be found e.g.
in~\cite{LindbladRodnianski};   the addition of the $u$ terms as in
\eq{Eins1} is straightforward).
Setting
$$
\Omega=\frac 1 {r^2}\;,\quad \tilde f = \Omega^{-\frac{n-2}2}f\;,
 \quad \tilde u= \Omega^{-\frac{n-2}2}u\;,\quad  y^i= \frac{x^i}{
 r^{2}}\;,
$$
one obtains a set of regular elliptic equations in the  coordinates
$y^i$ after a conformal rescaling $\delta_{ij}\to \Omega^2
\delta_{ij}$ of the flat metric, provided that $n\ge 6$. The reader
is referred to \cite{CCL} for a detailed calculation in a Lorentzian
setting, which carries over with minor modifications (due to the
quadratic rather than linear zero of $\Omega$)
to the current situation; note that $n$ in the calculations there
should be replaced by $n-1$ for the calculations at hand. We further
note that the leading order behavior of $\tilde f$ is governed by
the mass, which can be made arbitrarily small by a constant
rescaling of the metric and of the original harmonic coordinates
$x^i$; this freedom can be made use of to ensure ellipticity of the
resulting equations. Finally we emphasise  that this result,
contrary to the one for $n$ equal three or four, does not require
the non-vanishing of mass.

\section{Stationary vacuum solutions}
\label{Ssvs}

We consider Lorentzian metrics ${}^{n+1}g$ in odd
space-time-dimension $n+1\geq 7$, with Killing vector
$X=\partial/\partial t$. In adapted coordinates those metrics can be
written as
\beal{gme1}
 &^{n+1}g =
 -V^2(dt+\underbrace{\theta_i dx^i}_{=\theta})^2 +
 \underbrace{g_{ij}dx^i dx^j}_{=g}\;, & \\ &
 \partial_t V = \partial_t \theta = \partial_t g=0
 \;.
\eeal{gme2}
The vacuum Einstein equations (with vanishing cosmological constant)
read (see, e.g., \cite{Coquereaux:1988ne})
\begin{equation}\label{mainequation}
\left\{\begin{array}{l} V\nabla^*\nabla V=\frac 1{4} |\newF|_g^2\;,\\
     \Ric(g)-V^{-1}\Hess_gV=\frac{1}{2V^{2}}\newF\circ \newF\;,
\\
\divv  (V \newF)=0\;,
\end{array}\right.
\end{equation}
where
$$
 \newF_{ij}=-V^2(\partial_i \theta_j - \partial_j
 \theta_i)\;,\;\;\;(\newF\circ \newF)_{ij}=\newF_i{^k}\newF_{kj}\;.
$$
We consider metrics satisfying, for some $\alpha>0$,
\bel{foff}
  g_{ij}-\delta_{ij}=O(r^{-\alpha})\;, \ \partial_k
g_{ij}=O(r^{-\alpha-1})\;, \quad V= O(r^{-\alpha})\;,
 \ \partial_k V= O(r^{-\alpha-1})
 \;.
\ee
As is well known~\cite{Bartnik}, one can then introduce new
coordinates, compatible with the above fall-off requirements, which
are harmonic for $g$.

Next, a redefinition  $t\to t+ \psi$, introduces a gauge
transformation
$$
 \theta \to \theta + d\psi
 \;,
$$
and one can exploit this freedom to impose restrictions on $\theta$.
We will assume a condition of the form
\bel{thetagauge}
 g^{ij}\partial_i\theta_j= Q(\underbrace{g,V}_{p};\underbrace{\partial g, \partial V,  \theta}_{q})
 \;,
\ee
where $Q$ is a smooth function of the variables listed near
$(\delta,1;0,0,0)$, with a zero of order two or higher \emph{with
respect to $q$}:
$$
 Q(p;0)= \partial_q Q(p;0)=0
 \;.
$$
Examples include the \emph{harmonic gauge}, $\Box_{^{n+1}g}t=0$,
which reads
\bel{harmonic}
\partial_i(\sqrt{\det g} V g^{ij} \theta_j)=0
 \;,
\ee
as well as the maximal gauge,
\bel{maximal}
\partial_i(\frac {V^3 \sqrt{\det g}\, g^{ij}}{\sqrt{1-V^2g^{k\ell}\theta_k \theta_\ell}} \theta_j)=0
 \;.
\ee
Equation~\eq{harmonic} can always be achieved by solving a linear
equation for $\psi$, \emph{cf.,
e.g.},\/~\cite{choquet-bruhat:christodoulou:elliptic,Bartnik} for
the relevant isomorphism theorems. On the other hand, \eq{maximal}
can always be solved outside of some large ball~\cite{BCOM}. More
generally, when non-linear in $\theta$, equation \eq{thetagauge} can
typically be solved outside of some large ball using the implicit
function theorem in weighted H\"older or weighted Sobolev spaces.

In harmonic coordinates, and in a gauge \eq{thetagauge}, the system
\eq{mainequation} is elliptic and, similarly to the static case,
standard asymptotic considerations show that $g_{ij}$ is
Schwarzschild in the leading order, and that there exist constants
$\alpha_{ij}$ such that
$$
 \theta_i = \frac {\alpha_{ij}x^j}{r^n} + O(r^{-n})
 \;.
$$

To prove analyticity at $i^0$ one proceeds as in
Section~\ref{Sstatn}: thus, one first rewrites the second of
equations \eq{mainequation} as an equation for
$$
 \tilde g_{ij}:= e^{\frac{2u}{n-2}}g_{ij} \equiv V^{\frac 2 {n-2}}g_{ij}
 \;,
$$
which gets rid of the Hessian of $V$ there. It should then be clear
that, in coordinates which are harmonic for $\tilde g$, the first
two equations in \eq{mainequation} have the right structure for the
argument of Section~\ref{Sstatn}. It remains to check the third one.
For this we note that, in $\tilde g$--harmonic coordinates so that
$\partial_i(\sqrt{\det \tilde g}\, \tilde g^{ij})=0$,
\beaa
 \divv (V \lambda)_k
  &=&
   \frac1{\sqrt{\det g}}\partial_i\Big(\sqrt{\det g} V^3
   g^{ij}(\partial_j \theta_k - \partial_k \theta_j)\Big)
  \\
  &=&
   \frac{V^{\frac n {n-2}}}{\sqrt{\det \tilde g}}\partial_i\Big(\sqrt{\det \tilde g}
   V^2
   \tilde g^{ij}(\partial_j \theta_k - \partial_k \theta_j)\Big)
  \\
   &=&
   V^{\frac n {n-2}} \tilde g^{ij}\partial_i\Big( V^2
   (\partial_j \theta_k - \partial_k \theta_j)\Big) \\
 & = & V^{\frac n {n-2}} \Big(\tilde  g^{ij}\partial_i \partial_j \theta_k
      + 2V \tilde g^{ij}\partial_i V
   (\partial_j \theta_k - \partial_k \theta_j)
   \\
   &&
   - \underbrace{\tilde g^{ij} \partial_i \partial_k \theta_j}
   _{= V^{-\frac 2 {n-2}}(\partial_k(\underbrace{g^{ij}\partial_i \theta_j}_Q) + \partial_k g^{ij} \partial_i
   \theta_j)}\Big)
    \;.
\eeaa
If $Q$ in \eq{thetagauge} is zero, then the vanishing of $\divv
(V\lambda)$ immediately gives an equation of the right form for
$\theta$. Otherwise, $\partial Q$ leads to nonlinear terms of the
form $\partial^2_x g \,\theta$, \emph{etc}.,\/ which are again of
the right form, see  the calculations in~\cite{CCL}. Note that such
terms do not affect the ellipticity of the equations because of
their off-diagonal character.

\section{Einstein-Maxwell equations}
\label{SEMe}

The above considerations immediately generalise to the stationary
Einstein-Maxwell equations, with a Killing vector which approaches a
time-translation in the asymptotically flat region. Indeed, the
calculations of Section~\ref{Ssvs} carry over to this setting, as
follows:

Stationary Maxwell fields can be described by a  time-independent
scalar field $\varphi=A_0$ and a vector potential $A=A_idx^i$, again
time-independent. Here one needs to assume that, in addition to
\eq{foff}, one has
$$
A_\mu = O(r^{-\alpha})\;,\ \partial_k A_\mu = O(r^{-\alpha-1})
 \;.
$$
Maxwell fields lead to supplementary source terms in the
right-hand-sides of \eq{mainequation}
which are quadratic in the first derivatives of $\varphi$ and $A$,
hence of the right form for the argument so far. Next, if we write
the Maxwell equations as
$$
\frac 1 {\sqrt{-\det{}^{n+1}g}} \,\partial_\mu\Big( \sqrt{-\det{}
^{n+1}g}\ {}^{n+1}g^{\mu\rho}\
{}^{n+1}g^{\nu\sigma}\partial_{[\nu}A_{\sigma]}\Big)=0
 \;,
$$
and impose the Lorenz gauge,
$$
\frac 1 {\sqrt{-\det^{n+1}g}} \partial_\mu\Big( \sqrt{-\det{}
^{n+1}g}\ ^{n+1}g^{\mu\nu}A_{\nu}\Big)=0
 \;,
$$
the equations $\partial_t A_\mu=0$ allow one to rewrite the above as
$$
g^{ij}\partial_i\partial_j a = H(f,V,\theta;\partial f,\partial
V,\partial \theta;
\partial a )\;,
$$
where $a=(\varphi,A_i)$, with a function $H$ which is bilinear in
the second and third groups of arguments. This is again of the right
form, which finishes the proof of analyticity of $\tilde f,\tilde
\varphi,\tilde A$ and $\tilde\theta$ at $i_0$ for even $n\ge 6$,
where the original fields are related to the tilde-ones via a
rescaling by $\Omega^{\frac{n-2}2}$, \emph{e.g.}\/
$\varphi=\Omega^{\frac{n-2}2}\tilde \varphi$, and so on.

\bibliographystyle{amsplain}
%
\bibliography{../references/newbiblio,%
../references/reffile,%
../references/bibl,%
../references/hip_bib,%
../references/newbib,%
../references/PDE,%
../references/netbiblio}

%
%



\end{document}